\documentclass[reprint,aps,prb,amsmath,amssymb]{revtex4-1}

\usepackage[]{graphicx}
\usepackage[]{units}
\usepackage{times}
\usepackage{bm}
\usepackage{ulem}		
\usepackage{color}

\newcommand{\comment}[1]{}

\newcommand{\blue}[1]{{\color[rgb]{0,0,1}{#1}}} 

\renewcommand{\emph}{\textit}

\begin{document}

\title{Excitonic linewidth and coherence lifetime in monolayer transition metal dichalcogenides}
\author{Malte Selig$^1$}
\author{Gunnar Bergh\"auser$^1$}
\author{Archana Raja$^{2, 3}$}
\author{Philipp Nagler$^4$}
\author{Christian Sch\"uller$^4$}
\author{Tony F. Heinz$^{3, 5, 6}$}
\author{Tobias Korn$^4$}
\author{Alexey Chernikov$^{4, 6}$}
\author{Ermin Malic$^7$}
\author{Andreas Knorr$^1$}
\affiliation{$^1$Institut f\"ur Theoretische Physik, Technische Universit\"at Berlin,  10623 Berlin, Germany}
\affiliation{$^2$Department of Chemistry, Columbia University, New York, New York 10027, USA}
\affiliation{$^3$Department of Applied Physics, Stanford University, Stanford, California 94305, USA}
\affiliation{$^4$Institut f\"ur Experimentelle und Angewandte Physik, Universit\"at Regensburg, 93040 Regensburg, Germany}
\affiliation{$^5$SLAC National Accelerator Laboratory, Menlo Park, California 94025, USA}
\affiliation{$^6$Departments of Physics and Electrical Engineering, Columbia University, New York, New York 10027, USA}
\affiliation{$^7$Chalmers University of Technology, Department of Physics, SE-412 96 Gothenburg, Sweden}

\begin{abstract}

Atomically thin transition metal dichalcogenides (TMDs) are direct-gap semiconductors with strong light-matter and Coulomb interaction. The latter accounts for tightly bound excitons, which dominate the optical properties of these technologically promising materials. Besides the optically accessible bright excitons, these systems exhibit a variety of dark excitonic states. They are not visible in optical spectra, but can strongly influence the coherence lifetime and the linewidth of the emission from bright exciton states. In a recent study, an experimental evidence for the existence of such dark states has been demonstrated, as well as their strong impact on the quantum efficiency of light emission in TMDs.
Here, we reveal the microscopic origin of the excitonic coherence lifetime in two representative TMD materials (WS$_2$ and MoSe$_2$) within a joint study combining microscopic theory with optical experiments. We show that the excitonic coherence lifetime is determined by phonon-induced intra- and intervalley scattering into dark excitonic states. Remarkably, and in accordance with the theoretical prediction, we find an efficient exciton relaxation in WS$_2$ through phonon emission at all temperatures.
~\\
~\\

\end{abstract}

\maketitle

\begin{figure}[t!]
  \begin{center}
     \includegraphics[width=\linewidth]{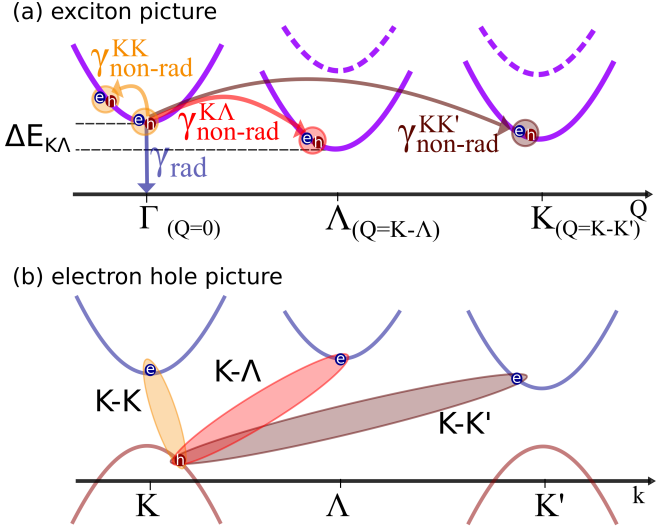}
   \end{center}
    \caption{\textbf{Relaxation channels determining the excitonic lifetime.} (a) Minima of the excitonic center of mass motion ($\mathbf{Q}$) dispersion $E(\mathbf{Q})$. An exciton at the $\Gamma$ point can decay via radiative $\gamma_{\text{rad}}$ (blue arrow) or non-radiative dephasing $\gamma_{\text{non-rad}}$. The latter occurs through exciton-phonon scattering within the $\Gamma$ valley (orange) or  to dark excitonic states at the $\Lambda$ (red) or the $K$ valley. 
For WS$_2$, the intervalley $\Lambda$ and $K$ excitons  lie energetically below the $\Gamma$ exciton ($\Delta E_{K \Lambda}<0$) allowing efficient scattering via emission of phonons even at very low temperatures.The dashed dispersion curves refer to a situation typical in MoSe$_2$, where $\Delta E_{K \Lambda}>0$. (b) An exciton at the $\Gamma$ ($\Lambda$,$K$) point is formed by an hole (red) at the $K$ point and an electron (blue) at the $K$ ($\Lambda$,$K'$) point.
}

  \label{scheme}
\end{figure}

As truly two-dimensional materials exhibiting a weak dielectric screening, monolayer transition metal dichalcogenides (TMDs) show a remarkably strong Coulomb interaction giving rise to the formation of tightly bound excitons.\cite{He2014,Chernikov2014,Berghauser2014,Ramasub2012} In addition to the optically accessible bright excitonic states located at the $K$ and $K^\prime$ points at the corners of the hexagonal Brillouin zone\cite{Mak2010,Splendiani2010,Radis2011,Lee2010}, there is also a variety of optically forbidden states including p excitons exhibiting a non-zero angular momentum, intravalley excitons with a non-zero center-of-mass momentum above the light cone as well as intervalley excitons, where a hole is located at the $K$ point and the electron either at the $K^\prime$ or the $\Lambda$ point,\cite{Louie2015,MacDonald2015,Steinhoff2015} cf. Fig. \ref{scheme}. In a recent time-resolved and temperature-dependent photoluminescence study, the existence of such dark intervalley excitons has been experimentally demonstrated.\cite{Zhang2015} In particular, it was shown that in tungsten-based TMDs, the intervalley dark exciton lies energetically below the optically accessible exciton resulting in a strong quenching of photoluminescence at low temperatures.\cite{Arora2015,Zhang2015}

Since excitons dominate the optical response of TMDs, a microscopic understanding of their properties is of crucial importance for their promising technological application in future optoelectronic and photonic devices. The presence of dark states has a strong impact on the coherence lifetime of optically accessible states, since they present a possible scattering channel that can be accessed via emission or absorption of phonons. The coherence lifetime in two\blue{-}dimensional TMDs is directly reflected by the homogeneous linewidth of excitonic resonances that is experimentally accessible in optical absorption and emission spectra.\cite{Kira2006} The homogeneous linewidth in monolayer WSe$_2$ has been recently measured via optical 2D Fourier transform spectroscopy allowing an unambiguous separation from inhomogeneous broadening.\cite{Moody2015} G. Moody et al. find, for temperatures up to \unit[50]{K}, a linear increase of the homogeneous linewidth in the range of \unit[4]{meV} to \unit[10]{meV}. Further studies show an increase of the linewidth up to \unit[40]{meV} in MoTe$_2$ and \unit[65]{meV} in MoS$_2$ at room temperature. \cite{Koirala2016, Dey2016} The observed increase is ascribed to scattering with acoustic phonons within the $K$ valley. Phonon-induced scattering into dark intervalley exciton states has not been considered. However, a consistent microscopic theory description of these observations is not available so far.

Here, we present a joint theory-experiment study aiming at a fundamental understanding of microscopic processes determining the excitonic coherence lifetime in TMDs. In the experiment, we use optical spectroscopy to extract temperature-dependent homogeneous linewidths from the total broadening of the exciton resonances in TMD monolayers.
Our theoretical approach is based on the Bloch equation for the microscopic polarization combined with the Wannier equation providing access to eigenvalues and eigenfunctions for bright and dark excitons. The joint study reveals the qualitatively different microscopic channels behind the excitonic coherence lifetime in tungsten- and molybdenum-based TMDs: In MoSe$_2$, the coherence lifetime is determined by radiative coupling at low temperatures and by phonon-induced intravalley scattering at room temperature. In contrast, the excitons in WS$_2$ can be efficiently scattered into the energetically lower\blue{-}lying intervalley dark excitonic state, cf. Fig. \ref{scheme}. This process is driven by phonon emission that is very efficient even at low temperatures. 
Note, that we focus on spin-conserving processes, occurring on an ultrashort time scale (tens of femtoseconds) in this study. Intervalley scattering via exchange interaction occurs on a much longer time scale (picoseconds\cite{Glazov2014,Song2013}) for supported samples.
To describe the coherence lifetime of optically bright excitons, we develop a theoretical model including all relevant relaxation channels on a microscopic footing. 
Furthermore, to test the theory, we perform experiments on WS$_2$ and MoSe$_2$ monolayers using a combination of linear reflectance and photoluminescence spectroscopy. 
From the obtained total linewidths of the exciton transitions we estimate temperature-dependent homogeneous broadening of the resonances corresponding to the coherence lifetime of the excitons.
Our analysis is also consistent with recent reports on total exciton linewidths in MoSe$_2$\,\cite{Arora2015}, as well as with the behavior of a similar TMD material WSe$_2$ studied through coherent spectroscopy at low-temperatures\,\cite{Moody2015} and under resonant excitation conditions\,\cite{Poellmann2015} (Additional details for the experimental procedure and data analysis are given in the Methods section).

The first step of the theoretical evaluation is the solution of the Wannier equation \cite{Berghauser2014,Axt2004, Kochbuch,Kira2006}
\begin{equation}
\frac{\hbar^2 \mathbf{q}^2 }{2m_0} \varphi_{\mathbf{q}}^{\mu}-\sum_{\mathbf{k}}V^{exc}_\mathbf{q,k} \varphi_{\mathbf{q+k}}^{\mu}+E_{gap} \varphi_{\mathbf{q}}^{\mu}=E^{\mu} \varphi_{\mathbf{q}}^{\mu},\label{wannier}
\end{equation} 
presenting an eigenvalue equation for excitons in TMDs. It includes the excitonic part of the Coulomb interaction $V^{exc}_\mathbf{q,k}$ that is treated within the Keldysh formalism for 2D systems.\cite{Berghauser2014, Cudazzo2011}
We obtain excitonic eigenenergies $E^{\mu}$ and excitonic wavefunctions 
$\varphi_{\mathbf{q}}^{\mu}$ for optically allowed bright and optically forbidden dark excitons that are denoted by the quantum number $\mu$. 
 The wavefunctions depend on the momentum $\mathbf{q}=\alpha \mathbf{k}_1+\beta \mathbf{k}_2$  describing the relative motion of electrons ($\bf k_1$) and holes ($\bf k_2$) in real space, where $\alpha = \frac{m_{e}}{m_{h}+m_{e}}$ and $\beta = \frac{m_{h}}{m_{h}+m_{e}}$ with effective masses for electrons and holes $m_e, m_h$.

The second step is to derive a Bloch equation for the 
microscopic polarization $p^{vc}_{\bf k_1 \bf k_2}(t)$ that determines the optical response of the material. The quantity reads in the excitonic basis\cite{Thranhardt2000} 
$
P^{\mu}_\mathbf{Q}=\sum_{\mathbf{q}} \varphi^{* \mu}_\mathbf{q} \langle a^{\dagger v}_\mathbf{q+\beta Q} a^{c}_\mathbf{q-\alpha Q} \rangle$
with $a^{(\dagger)\lambda}_{\mathbf{q}}$ as annihilation (creation) operators for an electron in the state $(\mathbf q, \lambda)$ with the band index $\lambda$ and $\mu$ being the quantum number of the exciton state. Here, we also have introduced the in-plane momentum $\mathbf{Q}=\mathbf{k}_1-\mathbf{k}_2$ denoting the Fourier coordinate of the center-of-mass motion in real space. 
The electronic dispersion is assumed to be parabolic, which is a good approximation in the vicinity of the $K$ point. This results in a quadratic dispersion for the excitons which constitutes the lowest excitonic contribution\cite{Louie2015,MacDonald2015} addressed in a coherent optical transmission experiment with fixed polarization.
We focus here on the energetically lowest intravalley and intervalley excitons numbered by $\mu$, i.e. the bound electron and hole are either both located at the $K$ point (intravalley), or only the hole is at the $K$ point, while the electron is either in the $K^\prime$ or in the $\Lambda$ point in the first Brillouin zone, cf. Fig. \ref{scheme}. Since a photon has a negligibly small center-of mass momentum $\bf Q$, only excitons with $\bf Q \approx 0$ are optically accessible. As a result, all intervalley excitons are dark.

Applying Heisenberg's equation of motion
$
i \hbar \partial_t \mathcal{P^{\mu}_\mathbf{Q}}=[H,\mathcal{P^{\mu}_\mathbf{Q}}],
$ 
we can determine the temporal evolution of the microscopic polarization $P^{\mu}_\mathbf{Q}$.
The Hamilton operator $
H$
includes (i) an interaction-free part containing the dispersion of electrons and phonons,
 (ii) the carrier-light interaction determining the optical selection rules,  (iii) the carrier-carrier interaction that has already been considered in the Wannier equation, 
and (iv) the carrier-phonon interaction coupling bright and dark excitons via emission and absorption of phonons.
The carrier-light coupling is considered within the semi-classical approach  in $\mathbf{p}\cdot\mathbf{A}$ gauge \cite{Carbonbuch}. The coupling is determined by the optical matrix element $M^{\sigma_-}_\mathbf{q}$ projected to the right-handed circular polarized light that is required to excite excitons at the $K$ point.\cite{Cui2012,Cao2012,Berghauser2014,Xiaodong2014} The carrier-phonon matrix element $g_{\bf q}^{\lambda \alpha}$ is treated within an effective deformation potential approach for acoustic phonons and approximating the Fr\"ohlich interaction for optical phonons.\cite{Li2013,Jin2014}

Evaluating the commutator in the Heisenberg equation of motion, we obtain in second-order Born-Markov-approximation\cite{Thranhardt2000} the Bloch equation for the microscopic polarizations  $P_\mathbf{Q}^{\mu}(t)$
\begin{align}
i\hbar \partial_{t} P_{\mathbf{Q}}^{\mu}(t)&=\left(\frac{\hbar^2 \mathbf{Q}^2}{2M}+E^{\mu}\right)P_{\mathbf{Q}}^{\mu}+i\hbar\sum_{\mathbf{q}} \varphi^{* \mu}_{\mathbf{q}}\Omega^{cv}_{\mathbf{q}}\delta_{\mathbf{Q},0}\nonumber \\&-i\frac{\pi}{\hbar}\sum_\mathbf{q',\alpha \nu} G_\mathbf{Q,Q+q'}^{\mu \nu \alpha}  P^{\nu}_{\mathbf{Q}}.\label{Blochgleichung}
\end{align}
While the first term describes the oscillation of the excitonic polarization determined by the excitonic dispersion in $\mathbf{Q}$, the second  term stands for  the carrier-light interaction given by the Rabi frequency $\Omega^{cv}_{\mathbf{q}}=\frac{e_0}{m_0}M^{\sigma_-}_\mathbf{q} A^{\sigma_-}$ with the vector potential $A^{\sigma_-}$ and the electron charge and mass $e_0, m_0$. The third contribution in equation (\ref{Blochgleichung}) describes the exciton-phonon-interaction that is given by the function
\begin{align}
G_\mathbf{Q,Q+q'}^{\mu \nu \alpha}=\hbar\sum_{\pm, \rho} g^{\mu \rho \alpha}_\mathbf{q'} g^{\rho \nu \alpha}_\mathbf{-q'} \left(\frac{1}{2} \pm \frac{1}{2} +n_\mathbf{q'}^{\alpha}\right) \times \nonumber \\
 \times \delta{(E^{\rho}(\mathbf{Q+q'})-E^{\nu}(\mathbf{Q}) \pm  \hbar\omega^{\alpha}_\mathbf{\pm q'})}.\label{phonCoupling}
\end{align}
Here, $\hbar\omega^{\alpha}_{\mathbf{q'}}$ is the phonon energy and $n_\mathbf{q'}^{\alpha}$ the phonon occupation in the mode $\alpha$ corresponding to the Bose-Einstein distribution. The function contains scattering processes including phonon emission ($+$) and absorption ($-$). The exciton-phonon coupling $
g_\mathbf{q'}^{\mu \nu \alpha}=\sum_{\mathbf{q}}( \varphi_{\mathbf{q}}^{*\mu} g_{\mathbf{q'}}^{c \alpha}  \varphi_{\mathbf{q-\beta q'}}^{\nu}-\varphi_{\mathbf{q}}^{*\mu} g_{\mathbf{q'}}^{v \alpha}  \varphi_{\mathbf{q+\alpha q'}}^{\nu})
$ depends on the electron-phonon matrix element $g_{\mathbf{q'}}^{c \alpha} $ and the overlap of the involved exciton wavefunctions in momentum space. The corresponding phonon-induced homogeneous dephasing of the excitonic polarization reads\cite{Thranhardt2000}
\begin{equation}
\gamma_\mathbf{Q}^{\mu \alpha}=\frac{\pi}{\hbar^{2}}\sum_\mathbf{q' \nu} G_\mathbf{Q,Q+q'}^{\mu \nu \alpha},
\label{gammaExcPh}
\end{equation}
giving rise to a non-radiative coherence lifetime for excitons with momentum $\mathbf{Q}$ in the state $\mu$.
Here, we take into account acoustic (LA, TA) and optical phonons (LO, TO),\cite{Jin2014} explicitly considering intravalley scattering between bright ($\bf Q=0$) and dark $K-K$ excitonic states ($\bf Q\neq0$) as well as intervalley scattering involving dark $K-\Lambda$ and $K-K'$ excitons, cf. Fig. \ref{scheme}. To evaluate the exciton-phonon scattering rates, we calculate the rate self-consistently, which corresponds to a self-consistent Born approximation.\cite{Schilp1994,Jauho2008}

Besides the exciton-phonon scattering, the coherence lifetime of excitons is also influenced by radiative coupling, i.e. spontaneous emission of light through recombination of electrons and holes. The radiative coupling is obtained by self-consistently solving the Bloch equation for the excitonic polarization and the Maxwell equations in a two dimensional geometry for the vector potential $A^{\sigma_-}$ yielding\cite{Knorr1996}
\begin{equation}
\gamma_{rad}=\frac{\hbar^2 c \mu_0 }{\omega n}|\sum_{\mathbf{q}} M^{\sigma_- *}_{\mathbf{q}} \varphi^{\mu}_{\mathbf{q}}|^{2}.\label{gammaRad}
\end{equation}
Here, $c/n$ is the light velocity in the substrate material and $\mu_0$ the vacuum permeability.

\begin{figure}[t!]
 \begin{center}
\includegraphics[width=0.95\linewidth]{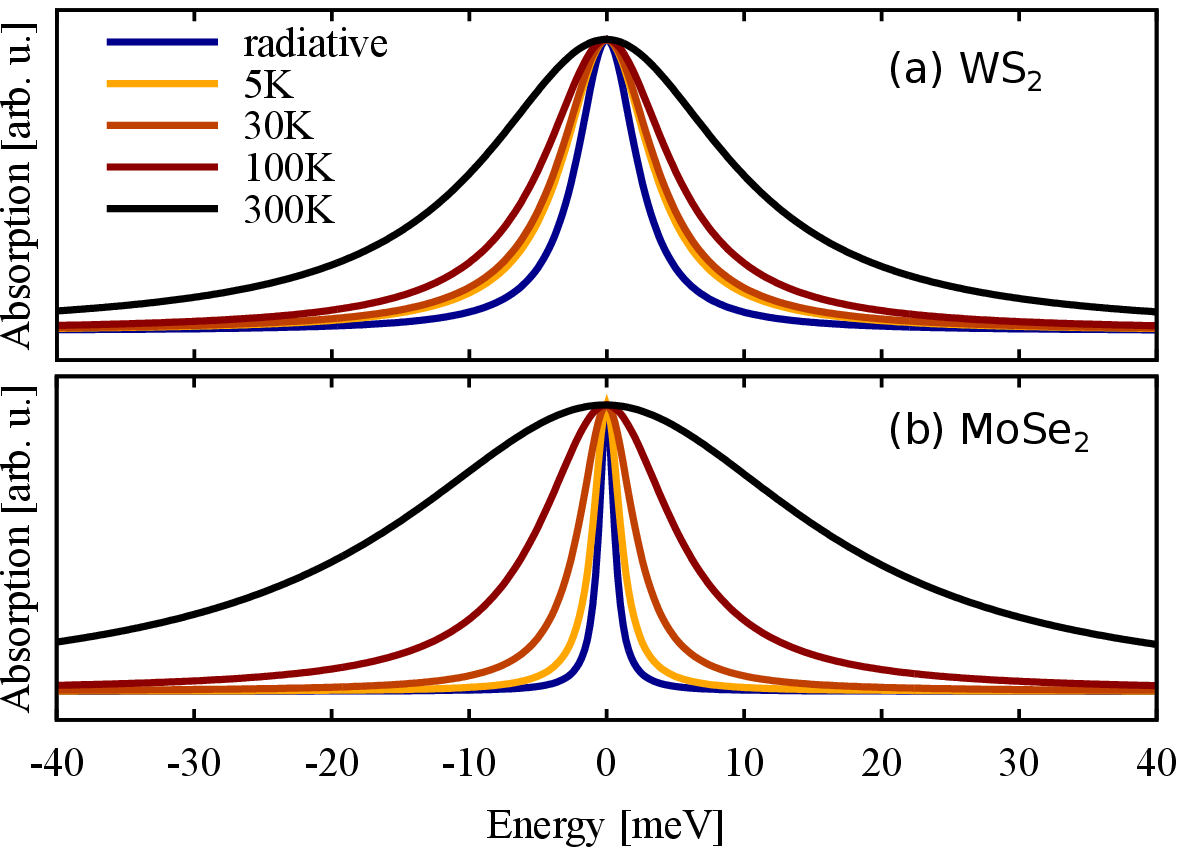}
\includegraphics[width=0.95\linewidth]{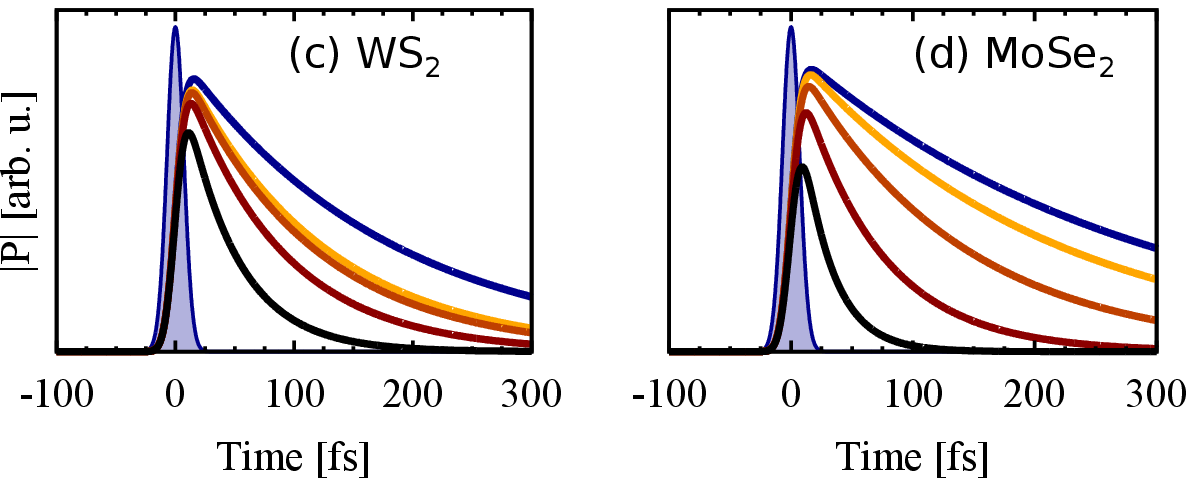}
 \end{center}
 \caption{\textbf{Homogeneous broadening and coherence lifetime.} Absorption spectrum of (a) WS$_2$ and (b) MoSe$_2$ focusing on the energetically lowest resonance of the A exciton. All spectra were normalized to their peak maximum. While the blue line only includes the radiative linewidth, the other lines contain also non-radiative contributions due to exciton-phonon scattering at different temperatures. Absolute value of the microscopic polarization for (c) WS$_2$ and (d) MoSe$_2$. The filled curve illustrates a \unit[10]{fs} excitation pulse at $t=0$.}
 \label{spectrum}
\end{figure}
\begin{figure}[t!]
  \begin{center}
    \includegraphics[width=0.95\linewidth]{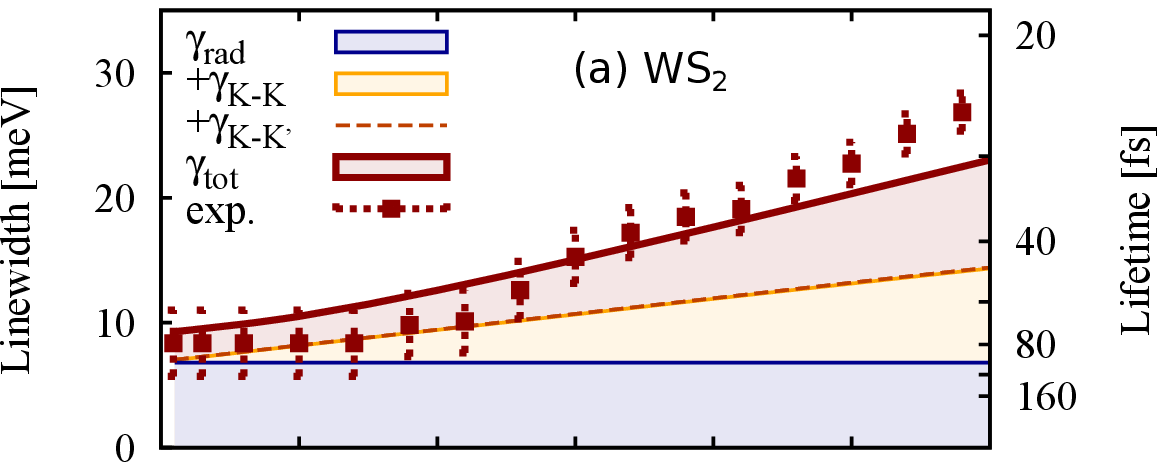}
    \includegraphics[width=0.95\linewidth]{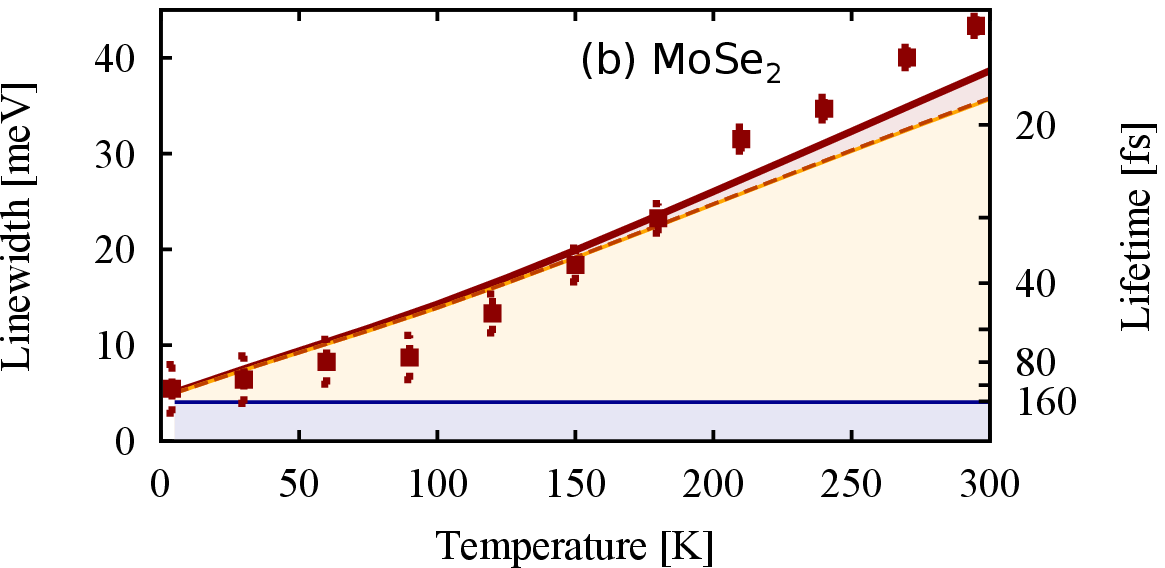}

  \end{center}
  \caption{\textbf{Excitonic linewidth and lifetime}. Temperature dependence of the linewidth and lifetime of the A exciton in (a) WS$_2$ and (b) MoSe$_2$ in dependence of temperature. The red points with error bars describe the experimental data. The thick red line shows the total linewidth consisting of the single contributions from the radiative decay $\gamma_{\text{rad}}$ (blue) and non-radiative decay including intravalley exciton-phonon coupling $\gamma_{\text{non-rad}}^{KK}$ (orange) as well as intervalley coupling $\gamma_{\text{non-rad}}^{K\Lambda}$ (only shown in the total contribution) and $\gamma_{\text{non-rad}}^{KK^\prime}$ (dashed orange). Note that the latter contribution is very small.}
 \label{gamma_temp}
\end{figure}

Having solved the Wannier equation, equation (\ref{wannier}), and the Bloch equation, equation (\ref{Blochgleichung}), we have access to the optical response of TMDs and can evaluate the radiative and non-radiative homogeneous linewidth of excitonic resonances reflecting the coherence lifetime of optically allowed excitons. Figure \ref{spectrum} illustrates the absorption spectrum of two dimensional sheets \cite{Knorr1996} of two exemplary monolayer TMDs including tungsten diselenide  (WS$_2$) and molybdenum diselenide (MoSe$_2$). We predict the homogeneous linewidth of the energetically lowest A exciton to be in the range of a few meV corresponding to an excitonic coherence lifetime of a few hundreds of femtoseconds. Depending on the  temperature, either the radiative or the non-radiative contribution is the dominant mechanism. 
We observe a larger radiative broadening in WS$_2$ (\unit[7]{meV} vs. \unit[4]{meV} for MoSe$_2$), while the overall broadening at room temperature is larger for MoSe$_2$ with  \unit[40]{meV} vs \unit[24]{meV} in WS$_2$.
Furthermore figure \ref{spectrum} shows the temporal evolution of the excitonic polarization $P_\mathbf{0}^{1s}(t)$ after optical excitation with a \unit[10]{fs} pulse. We find that the polarization decays radiatively with a time constant in the range of hundreds of fs. Including exciton-phonon coupling the time constant decreases drastically to some tens of fs at room temperature.
Evaluating equations (\ref{gammaExcPh}) and (\ref{gammaRad}), we can reveal the microscopic origin of the excitonic linewidth.  Figure \ref{gamma_temp} shows the temperature dependence of the linewidth depending on the nature of the coupling mechanism. Further, we show the corresponding exciton coherence lifetime which is connected by $L\ \tau = \hbar$, with $L$ being the full linewidth. We find an excellent agreement between theory and experiment with respect to qualitative trends as well as quantitative values for the linewidths.

In both investigated TMD materials, we observe a temperature-independent offset originating from radiative recombination. In contrast, the non-radiative coupling via scattering with phonons introduces a strong temperature dependence. Furthermore, we find a remarkably different behavior for MoSe$_2$ and WS$_2$: While for MoSe$_2$ intravalley exciton-phonon scattering is the crucial mechanism, the excitonic coherence lifetime in WS$_2$ is dominated by intervalley scattering $ \gamma_{\text{non-rad}}^{K\Lambda}$, coupling the optically allowed $K-K$ exciton with the dark $K-\Lambda$ states. The reason lies in the relative energetic position of these excitons. In WS$_2$, the $K-\Lambda$ exciton is located approximately \unit[70]{meV} below the $K-K$ exciton. 
Hence, exciton relaxation through phonon emission is very efficient even at 0 K resulting in a non-radiative offset in the homogeneous linewidth at low temperatures, cf. Fig. \ref{gamma_temp}(a). 
The coupling to $K-K'$ excitons does not significantly contribute due to the weak electron-phonon coupling element.\cite{Jin2014}

The situation is entirely different in MoSe$_2$, where
 the dark $K-\Lambda$ exciton lies approximately \unit[100]{meV} above the bright $K-K$ exciton. Thus, only the less efficient absorption of phonons can take place. Since the energy of large-momentum acoustic phonons,required for scattering $K-K$ excitons into $K-\Lambda$ states, is only  \unit[15]{meV} in this material,\cite{Jin2014} the only contribution of intervalley scattering in  MoSe$_2$ stems from the absorption of optical $\Gamma$ phonons and becomes relevant for temperatures higher than 150K, cf. Fig. \ref{gamma_temp}(b). 
Our calculations show that the intravalley scattering with acoustic $\Gamma$ phonons is the crucial mechanism for the excitonic coherence lifetime in MoSe$_2$. We observe a linear increase of the linewidth with temperature. Since acoustic $\Gamma$ phonons have small energies, the Bose-Einstein distribution appearing in equation (\ref{phonCoupling}) can be linearized resulting in a dephasing rate $\gamma_{\text{non-rad, ac}}^{KK^\prime}=2\pi^2|g_{q_0}|^2 \frac{k T}{2\hbar M c^2_{ac}}$ exhibiting a linear dependence on the temperature $T$. The slope is given by the exciton mass $M$, the velocity of the acoustic phonons $c_{ac}$, and the exciton-phonon coupling element $g_{q_0}$ at the position $q_0$, where the delta distribution is fulfilled, cf. equation (\ref{phonCoupling}).

The super-linear increase of the linewidth with temperature observed for both MoSe$_2$ and WS$_2$ can be ascribed to the scattering with optical $\Gamma$ or $\Lambda$ phonons. 
The overall temperature dependence of the excitonic linewidth can be approximated by 
$
\gamma=\gamma_0+ c_1 T + \frac{c_2}{e^\frac{\Omega}{kT}-1}.
$
For WS$_2$, we find a temperature-independent offset of $\gamma_0$=\unit[9.1]{meV} consisting of \unit[7]{meV} due to radiative decay and \unit[2.1]{meV} due to acoustic $\Lambda$ phonon emission, furthermore the slope $c_1=\unit[28]{\frac{\mu \text{eV}}{K}}$ describing the linear increase due to acoustic $\Gamma$ phonons, the rate $c_2=\unit[6.5]{meV}$ and the averaged energy  $\Omega=\unit[20]{meV}$ of involved acoustic $\Gamma$ phonons determining the strength of the superlinear increase. We find that optical $\Gamma$ phonons do not give a contribution to the superlinear increase.
The corresponding parameters for MoSe$_2$ read $\gamma_0=\unit[4.3]{meV}$, $c_1=\unit[91]{\frac{\mu \text{eV}}{K}}$, $c_2=\unit[15.6]{meV}$ ($c_2=\unit[7.2]{meV}$ due to intravalley optical phonon scattering and $c_2=\unit[8.4]{meV}$ due to $K-\Lambda$ coupling), and $\Omega=\unit[30]{meV}$. Our results for the radiative dephasing are in good agreement with recent calculations.\cite{Wang2016}

In conclusion, we have presented a joint theory-experiment study revealing the microscopic origin of the excitonic lifetime in atomically thin 2D materials.
We find both in theory and experiment a qualitatively different origin of the coherence lifetime limiting processes in tungsten- and molybdenum-based TMDs. While in MoSe$_2$, the coherence lifetime of an optically bright exciton is determined by intravalley scattering with acoustic phonons, in WS$_2$  scattering into dark excitonic $K-\Lambda$ states is crucial. The gained insights shed light into excitonic properties that are crucial for exploiting the technological potential of these atomically thin nanomaterials.
In particular, it will allow us to access exciton dynamics on a microscopic level across a large variety of relevant experimental scenarios, including exciton formation, thermalization and relaxation among many others.
The presented theoretical approach can be furthermore generalized to quantitatively describe exciton behavior for the whole family of semiconducting 2D materials beyond the representative systems studied here.
It provides a theoretical basis to explore fundamental many-body physics of 2D materials, crucial for future applications and allowing for consistent theoretical predictions of the functionality for novel devices on inter-atomic scales.

\section*{Acknowledgements}
We acknowledge financial support from the Deutsche Forschungsgemeinschaft (DFG) through SFB 951 (A.K.) SFB 787 (M.S.), SFB 689 (T.K. and C.S.), GK 1570 (P.N.), Emmy Noether Program (A.C.) and the EU Graphene Flagship (CNECT-ICT-604391)(E.M.,G.B.). This work was further supported by the U.S. Department of Energy, Office of Science, Office of Basic Energy Sciences, with funding at Columbia University through the Energy Frontier Research Center under grant DE-SC0001085 for optical measurements and at SLAC National Accelerator Laboratory through the AMOS program within the Chemical Sciences, Geosciences, and Biosciences Division for data analysis. A.C. gratefully acknowledges funding from the Keck Foundation.

~\\
~\\


%

\end{document}